\documentclass[twocolumn]{jpsj3} %% two-column layout
\title{
Electronic Structure of Heavy Fermion Uranium Compounds Studied by Core-Level Photoelectron Spectroscopy
}

\author{
Shin-ichi~\textsc{Fujimori}$^1$\thanks{E-mail address: fujimori@spring8.or.jp},
Takuo~\textsc{Ohkochi}$^{1}$\thanks{Present address: Japan Synchrotron Radiation Research Institute, Sayo, Hyogo 679-5198, Japan},
Ikuto~\textsc{Kawasaki}$^1$,
Akira~\textsc{Yasui}$^1$,
Yukiharu~\textsc{Takeda}$^1$,
Tetsuo~\textsc{Okane}$^1$,
Yuji~\textsc{Saitoh}$^1$,
Atsushi~\textsc{Fujimori}$^{1,2}$,
Hiroshi~\textsc{Yamagami}$^{1,3}$,
Yoshinori~\textsc{Haga}$^{4,5}$,
Etsuji~\textsc{Yamamoto}$^{4}$,
Yoshifumi~\textsc{Tokiwa}$^{4}$\thanks{Present address: Physikalisches Institut, Georg-August-Universitat G\"ottingen, D-37077 G\"ottingen, Germany},
Shugo~\textsc{Ikeda}$^{4,6}$\thanks{Present address: Graduate School of Material Science, University of Hyogo, Kamigori, Hyogo 678-1297, Japan},
Takashi~\textsc{Sugai}$^{4,5}$,
Hitoshi~\textsc{Ohkuni}$^{6}$,
Noriaki~\textsc{Kimura}$^{6}$\thanks{Present address: Graduate School of Science, Tohoku University, Sendai 980-8578, Japan},
and Yoshichika~\textsc{\=Onuki}$^{4,6}$
}

\inst{
$^{1}$Condensed Matter Science Division, Japan Atomic Energy Agency, Sayo, Hyogo 679-5148, Japan\\
$^{2}$Department of Physics, University of Tokyo, Hongo, Tokyo 113-0033, Japan\\
$^{3}$Department of Physics, Faculty of Science, Kyoto Sangyo University, Kyoto 603-8555, Japan\\
$^{4}$Advanced Science Research Center, Japan Atomic Energy Agency, Tokai, Ibaraki 319-1195, Japan\\
$^{5}$Graduate School of Science, Tohoku University, Sendai 980-8578, Japan\\
$^{6}$Graduate School of Science, Osaka University, Toyonaka, Osaka 560-0043, Japan
}

\abst{
High-energy-resolution core-level and valence-band photoelectron spectroscopic studies were performed for the heavy Fermion uranium compounds UGe$_2$, UCoGe, URhGe, URu$_2$Si$_2$, UNi$_2$Al$_3$, UPd$_2$Al$_3$, and UPt$_3$ as well as typical localized and itinerant uranium compounds to understand the relationship between the uranium valence state and their core-level spectral line shapes.
In addition to the main line and high-binding energy satellite structure recognized in the core-level spectra of uranium compounds, a shoulder structure on the lower binding energy side of the main lines of localized and nearly localized uranium compounds was also found.
The spectral line shapes show a systematic variation depending on the U~5$f$ electronic structure.
The core-level spectra of UGe$_2$, UCoGe, URhGe, URu$_2$Si$_2$, and UNi$_2$Al$_3$ are rather similar to those of itinerant compounds, suggesting that U~5$f$ electrons in these compounds are well hybridized with ligand states.
On the other hand, the core-level spectra of UPd$_2$Al$_3$ and UPt$_3$ show considerably different spectral line shapes from those of the itinerant compounds, suggesting that U~5$f$ electrons in UPd$_2$Al$_3$ and UPt$_3$ are less hybridized with ligand states, leading to the correlated nature of U~5$f$ electrons in these compounds.
The dominant final state characters in their core-level spectra suggest that the numbers of 5$f$ electrons in UGe$_2$, UCoGe, URhGe, URu$_2$Si$_2$, UNi$_2$Al$_3$, and UPd$_2$Al$_3$ are close to but less than three, while that of UPt$_3$ is close to two rather than to three.
}

\kword{photoelectron spectroscopy, uranium compounds, heavy fermion, electronic structure}

\begin{document}
\maketitle
%-----------------------------------------------------------------------------------------
\section{Introduction} 
Heavy Fermion uranium compounds show a wide variety of physical properties such as unconventional superconductivity, various magnetic orderings, and their coexistence.
To understand the origin of these physical properties, the valence states of the uranium atom, namely, the number of 5$f$ electrons and how they are hybridized with the ligand states, are the most essential pieces of information for modeling their electronic structures. 
However, the valence states of the uranium atom, especially in metallic compounds, are generally unknown owing to the lack of appropriate experimental methods of determining them.
In the present study, we have explored the possibility of determining the valence states of the uranium atoms by core-level spectroscopy.

Core-level spectroscopy is a very powerful experimental technique for studying the valence states of atoms in solids\cite{core}.
The different valence states manifest themselves as the binding energy shifts of core-level spectra due to the different degrees of screening of the nuclear potential by valence electrons.
This is called a chemical shift, and has been used to identify the valence state of atoms in solids as well as of molecules.
%-----------------------------------------------------------------------------------------
\begin{figure}
\begin{center}
\includegraphics[scale=0.5]{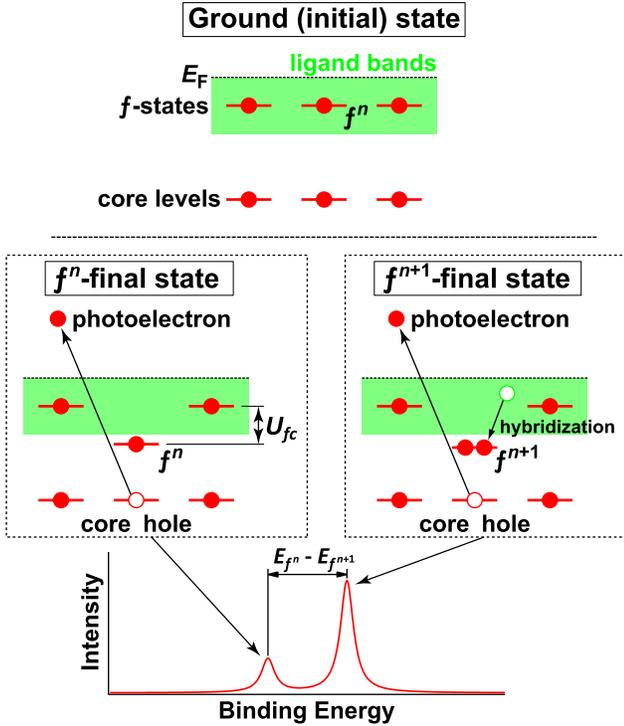}
\caption{(Color-online) Origin of multiple final states in strongly correlated $f$-electron materials.
In the final state, the energy of $f$ states in the core hole site is lowered by its attractive potential $U_{fc}$.
The screening of the core hole potential may occur by occupying additional $f$ states through charge transfer from the ligand states depending on the local electronic structure.
In general, the $f^{n+1}$ final state becomes dominant in compounds with strong hybridization.
Note that if a material is in a mixed valence state, three or more peaks should appear in the core-level spectrum.
}
\label{coreprocess}
\end{center}
\end{figure}
%-----------------------------------------------------------------------------------------
In addition to the chemical shift, there are contributions of final-state effects.
For example, the core-level spectra of strongly correlated materials are accompanied by complex satellite structures due to the existence of multiple final states in the photoelectron excitation process.
In Fig.~\ref{coreprocess}, the origin of multiple final-state peaks in the core-level spectra of strongly correlated $f$ electron materials is shown schematically.
In the final state of core-electron emission, there is a core hole that lowers the energy of the local $f$ states.
If there is considerable hybridization between $f$ and ligand states, and the energy of the screened final state ($f^{n+1}$+core hole) is lower than that of the nonscreened state ($f^{n}$+core hole), the screened final-state peak becomes dominant in photoemission spectra.
Here, note that the final state with a lower energy corresponds to a peak on the low-binding-energy side.
In this way, the core-level spectra of strongly correlated $f$ electron compounds may have multiple final states depending on their electronic structures.
If the material is in a mixed valence state with the $f^n$ and $f^{n+1}$ configurations, three peaks representing $f^n$, $f^{n+1}$, and $f^{n+2}$ final-state configurations can appear in the core-level spectrum.
For cerium-based materials, their core-level spectra consist of well-separated three-peak structure representing the 4$f^0$, 4$f^1$, and 4$f^2$ final states\cite{Ce3d}.
Their relative energy positions and intensities are different depending on the compound, and they have been analyzed using the single-impurity Anderson model\cite{GS}.
Basic physical parameters such as the bare $f$ electron energy $\epsilon_f$, the Coulomb potential $U_{ff}$, the hybridization strength $V_{fc}$, and the number of $f$ electrons in the ground state can be derived from their analysis.
Therefore, the number of $f$ electrons $n_f$ can be determined by analyzing their core-level spectra.

The core-level spectra of uranium compounds also show multiple-final-state peak structures, and their shapes vary widely variations from compound to compound\cite{Ucore}.
There are a number of studies of their core-level spectra, but the microscopic origin of the spectral line shape is not well understood yet.
Fujimori {\it et al.} studied the core-level spectra of some heavy-Fermion uranium compounds as well as their diluted alloys, and suggested that their core-level spectral line shapes are essentially governed by the local electronic structure around uranium sites\cite{UM2Al3_XPS}.
This implies that an impurity model similar to the model that has been used for Ce-based compounds is applicable to uranium compounds as well.
However, their peak structures are relatively broad compared with those of Ce-based compounds, and it is very difficult to identify the contributions of each different final state.
Since the overall peak structure consists of a relatively sharp and dominant main line on the low-binding-energy side and a relatively broad and weak satellite on the high-binding-energy side, the core-level spectra of uranium compounds have been analyzed by assuming two different final states.
Ejima {\it et al.}\cite{Ejima} analyzed the core-level spectra of various uranium compounds with two different final states.
They have interpreted their results with the Kotani-Toyozawa model\cite{KT1,KT2}, which has been used to describe the core-level spectra of La-based compounds.
On the other hand, systematic photoemission studies of uranium compounds\cite{UM2Al3_XPS} and alloys\cite{URhPd} suggested that there should be more than two peak structures, which however were not directly observed experimentally.
From a theoretical point of view, Okada pointed out that at least four final-state configurations ($f^2$, $f^3$, $f^4$, and $f^5$ final states) have finite contributions to the core-level spectral line shape\cite{Okada}.
However, a larger number of $f$ electrons ($n_f$=2-3) in the initial state, and the strong hybridization between $f$ and ligand states make the number of matrix elements too large to perform a realistic simulation of core-level spectra practically. 
Therefore, it remains to be understood at present, and the relationship between the valence state and the core-level spectral line shape has not been well understood yet, although there have been some attempts.

In the present study, we have experimentally studied the core-level spectra of uranium compounds to understand the relationship between uranium valence states and core-level spectra.
The present paper is organized as follows.
First, we study the typical itinerant 5$f$ and localized 5$f$ compounds to understand the basic behavior of the core-level spectra of uranium compounds.
We take some itinerant compounds to observe variations within the itinerant compounds.
Then, we study the core-level spectra of heavy-Fermion uranium superconductors where the valence states are controversial.
Finally, we compare all these spectra and try to understand their electronic structures as well as the valence states from their core-level spectral line shapes.

%-----------------------------------------------------------------------------------------
\section{Summary of Studied Compounds} 
%-----------------------------------------------------------------------------------------
\begin{table*}[htb]
\caption{Physical properties of uranium compounds studied in the present study.}
%\begin{ruledtabular}
\begin{tabular}{cccccc}
\hline
% & \raisebox {0.5zh}[0cm][0cm]{Structure} & \shortstack{$\gamma$ \\ {\tiny (mJ/molK$^2$)}} & \shortstack{$T_{\rm ord}$ \\ (K)} & \shortstack{$\mu_{\rm ord}$ \\ ($\mu_{\rm B}$)} & \shortstack{$T_{\rm SC}$ \\ (K)} \\
 & Structure & \shortstack{$\gamma$ \\ {\tiny (mJ/molK$^2$)}} & \shortstack{$T_{\rm ord}$ \\ (K)} & \shortstack{$\mu_{\rm ord}$ \\ ($\mu_{\rm B}$)} & \shortstack{$T_{\rm SC}$ \\ (K)} \\
\hline
\multicolumn{4}{l}{\bf localized system}\\
UPd$_3$ & hexagonal & 7.6 & & & \\
\hline
\multicolumn{4}{l}{\bf Pauli-paramagnet}\\
UB$_2$ & hexagonal & 10 & & & \\
UFeGa$_5$ & tetragonal & 37 & & & \\
\hline
\multicolumn{4}{l}{\bf antiferromagnet}\\
UPtGa$_5$ & tetragonal & 67 & 26 & 0.25 & \\
\hline
\multicolumn{4}{l}{\bf heavy Fermion superconductors}\\
UGe$_2$ & orthorhombic & 32 & 52 (F) & 1.48 & 0.8*\\
UCoGe & orthorhombic & 57 & 3 (F) & 0.07 & 0.8\\
URhGe & orthorhombic & 155 & 9.5 (F) & 0.42 & 0.25\\
URu$_2$Si$_2$ & tetragonal & 180 & 17.5 (HO$^\dagger$) &$<$0.02 & 1.4\\
UNi$_2$Al$_3$ & hexagonal & 145 & 4.6 (AF) & 0.24 & 1 \\
UPd$_2$Al$_3$ & hexagonal & 210 & 14 (AF) & 0.85 & 2\\
UPt$_3$ & hexagonal & 420 & 5 (AF) & 0.02 & 0.4 \\
%USb$_2$ & tetragonal & 26 & 203 & 1.9 & \\
%UN & cubic & 49 & 51 & 0.75 & \\
\hline
\multicolumn{4}{l}{*Under pressure of 1.2~GPa}\\{$^\dagger$Hidden-order transition}\\
\end{tabular}
%\end{ruledtabular}
\end{table*}
%-----------------------------------------------------------------------------------------
%
Table I shows a summary of the physical properties of the uranium compounds investigated in the present study.

%-----------------------------------------------------------------------------------------
%UPd3
%-----------------------------------------------------------------------------------------
UPd$_3$ is a localized uranium compound with a 5$f^2$ configuration\cite{UPd3}.
Its specific heat coefficient is as small as $\gamma$=7.6~mJ/molK$^2$ since U~5$f$ electrons are completely localized.
In previous angle-resolved photoemission (ARPES) experiments, it was shown that the nature of the Fermi surface is governed by non-$f$ states\cite{UPd3_Itoh}.

%-----------------------------------------------------------------------------------------
%UB2
%-----------------------------------------------------------------------------------------
UB$_2$ is an itinerant compound\cite{UB2}.
Its electronic specific heat coefficient is a very small since 5$f$ electrons have very itinerant character.
The U-U distance in this compound is 3.123 \AA, which is much smaller than the so-called Hill limit (3.4 \AA).
It has been suggested that the direct overlap of the U~5$f$ wave functions becomes dominant if the U-U distance is smaller than this value.
Therefore, the direct overlap of the U~5$f$ wave functions plays an essential role in its electronic structure.
An ARPES study of this compound showed that U~5$f$ electrons form itinerant bands near $E_{\rm F}$, and its band structure and Fermi surfaces are well explained by the band-structure calculation treating all U~5$f$ electrons as itinerant\cite{UB2_ARPES}.
Therefore, U~5$f$ electrons have a very itinerant nature in this compound.

%-----------------------------------------------------------------------------------------
%UTGa5
%-----------------------------------------------------------------------------------------
UFeGa$_5$ is also an itinerant paramagnetic compound\cite{UFeGa5_dhva}.
An ARPES study of this compound revealed that its band structure as well as Fermi surface are well explained by the band-structure calculation\cite{UFeGa5_ARPES}.
Meanwhile, its specific heat coefficient is considerably larger than that of UB$_2$, suggesting that the U~5$f$ electrons are not as itinerant as those in UB$_2$.
UPtGa$_5$ has the same crystal structure as UFeGa$_5$, but it shows antiferromagnetic transition at $T_{\rm N}$=26~K.
The dHvA experiment suggested the itinerant nature of 5$f$ states in this compound\cite{UPtGa5_dHvA}.
Since the U-U distances in these compounds are much larger than the Hill limit (4.258~\AA~for UFeGa$_5$ and 4.341~\AA~for UPtGa$_5$), the hybridization between U~5$f$ and ligand states governs their electronic structures.

%-----------------------------------------------------------------------------------------
%UGe2
%-----------------------------------------------------------------------------------------
UGe$_2$ is a ferromagnetic (F) compound with a Curie temperature of $T_{\rm C}$=52~K.
It undergoes a superconducting transition at $T_{\rm SC}$=0.8~K at a pressure of about 1.2~GPa\cite{UGe2_SC}.
This superconductivity coexists with ferromagnetic ordering.
The itinerant nature of the U~5$f$ states was suggested for UGe$_2$ from both dHvA experiments and transport investigations\cite{UGe2_dHvA}.

%-----------------------------------------------------------------------------------------
%UTGe
%-----------------------------------------------------------------------------------------
UCoGe \cite{UCoGe} and URhGe \cite{URhGe} are ferromagnetic compounds with a Curie temperatures of $T_{\rm C}$=3~K (UCoGe) and 9.5~K (URhGe).
Both of them exhibit $p$-type superconductivity below $T_{\rm SC}$=0.8~K (UCoGe) and 0.25~K (URhGe), and superconductivity coexists with ferromagnetic ordering.
The nature of the 5$f$ electrons in these compounds is considered to be itinerant, but their electronic structures are not well understood yet.

%-----------------------------------------------------------------------------------------
%URu2Si2
%-----------------------------------------------------------------------------------------
URu$_2$Si$_2$ is a heavy Fermion superconductor with a tetragonal crystal structure.
This compound has been well known as a compound that shows ''hidden-order'' transition at 17.5~K \cite{URu2Si2}.
It undergoes superconducting transition at $T_{\rm SC}$=1.4~K.
The nature of U~5$f$ electrons in this compound is still controversial, but the recent photoemission study showed that U~5$f$ electrons form quasi-particle bands near $E_{\rm F}$, suggesting that they have an itinerant nature\cite{URu2Si2_ARPES}.

%-----------------------------------------------------------------------------------------
%UT2Al3
%-----------------------------------------------------------------------------------------
UPd$_2$Al$_3$~\cite{UPd2Al3} and UNi$_2$Al$_3$~\cite{UNi2Al3} are also heavy Fermion superconductors.
They show the coexistence of unconventional superconductivity and long-range antiferromagnetic (AF) order with relatively large magnetic moments.
Both compounds have a common hexagonal crystal structure in which two-dimensional U and Pd or Ni layers and Al layers are stacked alternatively along the $c$-axis.
UNi$_2$Al$_3$ has smaller lattice constants than UPd$_2$Al$_3$, leading to more itinerant U~5$f$ properties.
This can be inferred from the smaller electronic specific heat coefficient of UNi$_2$Al$_3$ than of UPd$_2$Al$_3$.
An ARPES study of these compounds revealed that there exist quasi-particle bands originating from U~5$f$ states \cite{UPd2Al3_ARPES}.
Despite their similarities, their magnetic and superconducting properties are considerably different.
UPd$_2$Al$_3$ exhibits a simple antiferromagnetic structure with a large static magnetic moment (0.85~$\mu_{\rm B}$) lying in the basal plane.
Owing to the large local magnetic moment and the large entropy release $\Delta S$ at $T_{\rm N}$, it has been argued that this magnetic ordering originates from localized U~5$f$ states.
On the other hand, the magnetic structure of UNi$_2$Al$_3$ shows incommensurate SDW-type ordering.
Therefore, different mechanisms have been proposed for the origin of magnetism in these compounds.
The symmetry of Cooper pairing is singlet in UPd$_2$Al$_3$, although it has been suggested to be triplet in UNi$_2$Al$_3$.

%-----------------------------------------------------------------------------------------
%UPt3
%-----------------------------------------------------------------------------------------
UPt$_3$ is a heavy Fermion superconductor with a very large specific heat coefficient of $\gamma$=420 mJ/molK$^2$\cite{UPt3}.
Recent dHvA experiments have suggested that band-structure calculation with fully itinerant U~5$f$ electrons gives a rather good description of the observed Fermi surfaces\cite{UPt3_dHvA}.
However, this extremely enhanced specific heat coefficient implies that U~5$f$ electrons are located at the borderline between localized and itinerant states.

\section{Experimental Procedure}
Photoemission experiments were performed at the soft X-ray beamline BL23SU of SPring-8~\cite{BL23SU} using a photoemission spectrometer equipped with a Gammadata-Scienta SES-2002 electron analyzer.
The energy resolution was set to be 100-150~meV at $h\nu$=800~eV.
In this photon energy, contributions from the U~5$f$ and transition metal $d$ states are dominant in the valence-band spectra\cite{Lindau}.
The position of the Fermi level was determined by that of an {\it in situ}-evaporated gold film.
All measurements were carried out at 20~K.
Therefore, the measured compounds were in the paramagnetic phase except for UPtGa$_5$ and UGe$_2$, which were in the antiferromagnetic and ferromagnetic phases, respectively.
All the samples were single crystals, and the clean sample surfaces were obtained by cleaving under an ultrahigh-vacuum condition.
The Shirley-type inelastic background\cite{Shirley} was removed in both core-level and valence-band spectra.

\section{Results and Discussion}

\subsection{Itinerant and localized 5$f$ compounds}
%-----------------------------------------------------------------------------------------
\begin{figure}
\begin{center}
\includegraphics[scale=0.45]{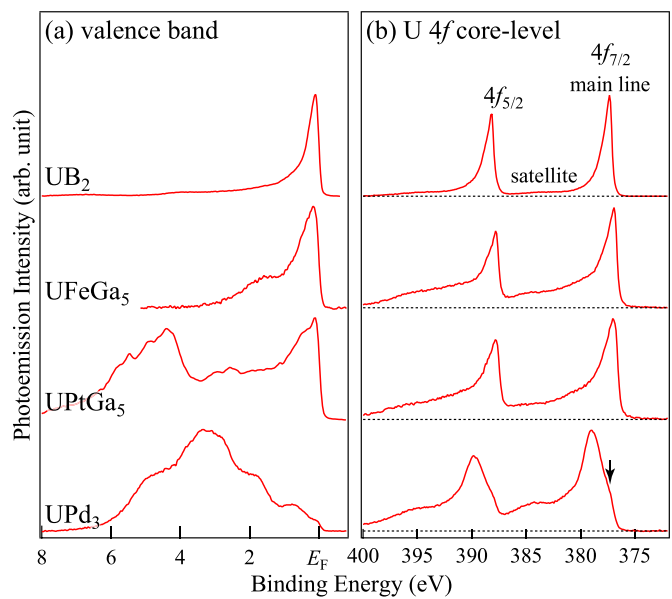}
\caption{(Color-online)
Valence-band spectra (a) and U~4$f$ core-level spectra (b) of UB$_2$, UFeGa$_5$, UPtGa$_5$, and UPd$_3$ taken at $h\nu=800$~eV.
The horizontal lines in the bottom of the spectra show the zero lines after background subtraction.
The position of the low-binding-energy shoulder of UPd$_3$ is indicated by the arrow.}
\label{UB2_UTGa5_UPd3}
\end{center}
\end{figure}
%-----------------------------------------------------------------------------------------
First, we discuss the valence-band and U~4$f$ core-level spectra of the typical itinerant 5$f$ compounds UB$_2$, UFeGa$_5$, and UPtGa$_5$ and the localized 5$f$ compound UPd$_3$.
Figure~\ref{UB2_UTGa5_UPd3}(a) shows their valence-band spectra.
In the valence-band spectrum of UB$_2$, there is a dominant sharp peak just below $E_{\rm F}$.
They are contributions mainly of U~5$f$-originated bands.
In previous ARPES studies of UB$_2$, clear energy band dispersions with an energy order of 1~eV were observed in this energy region\cite{UB2_ARPES}.
This is consistent with the very itinerant nature of U~5$f$ states in this compound.
In the valence-band spectrum of UFeGa$_5$, there is a predominant asymmetric peak in the vicinity of $E_{\rm F}$.
According to our previous ARPES study of UFeGa$_5$, this is the contribution of the broad U~5$f$ bands with an energy dispersion of about 0.5~eV\cite{UFeGa5_ARPES}.
The Fe~3$d$ bands are located at approximately $E_{\rm B}$=1.5~eV.
Meanwhile, in the valence-band spectrum of UPtGa$_5$, a similar predominant asymmetric peak was observed just below $E_{\rm F}$.
In addition, there is a shoulder peak structure at approximately $E_{\rm B}$=0.5~eV.
Since the intensity of this shoulder structure has the same photon energy dependence of the peak just below $E_{\rm F}$, this is the contribution mainly of U~5$f$ states.
The band structure calculation on UPtGa$_5$ in the antiferromagnetic phase shows that partial U~5$f$ density states have similar structures around this energy range.
This suggests that this shoulder structure originates from the band structure of UPtGa$_5$.
On the higher-binding-energy side ($E_{\rm B}$=3-7~eV), there are strong and broad peak structures.
These contribution originates mainly from the U~5$f$ and Pt~5$d$ states, respectively.
The peak near $E_{\rm F}$ is narrower than that of UFeGa$_5$, and this is consistent with the larger specific heat coefficient of UPtGa$_5$ than of UFeGa$_5$.

In contrast to the itinerant compounds, UPd$_3$ showed no sharp peak structure just below $E_{\rm F}$ in its valence-band spectrum.
The density of states at $E_{\rm F}$ is very low, suggesting the absence of U~5$f$-originated bands in the vicinity of $E_{\rm F}$. 
Instead, the contributions of Pd~4$d$ states are distributed in a wide energy range from $E_{\rm B}$=6~eV to $E_{\rm F}$, suggesting that the Fermi surface character is dominated by Pd~4$d$ states.
This is consistent with the localized nature of U~5$f$ states in this compound.
Although the contribution of U~5$f$ states is not clearly distinguished in the present spectrum, it should be distributed in the energy range of $E_{\rm B}$=0.5-1~eV, according to previous resonant photoemission experiments on this compound\cite{UPd3_RPES}.

Next, we show the core-level spectra of those four compounds.
Figure~\ref{UB2_UTGa5_UPd3}(b) shows the U~4$f$ core-level spectra of UB$_2$, UFeGa$_5$, UPtGa$_5$, and UPd$_3$.
The U~4$f$ core levels have been chosen for the present study because they have a large photoemission cross section in the soft-X-ray region, and the lifetime broadening is small enough to observe their fine structures.
In addition, their spin-orbit splitting energy is as large as about 12~eV corresponding to the U~4$f_{7/2}$ and U~4$f_{5/2}$ components, and this energy separation makes it possible to distinguish the contributions of each spin-orbit-split component. 
Both components show essentially the same peak structure.
It is shown that the U~4$f$ core-level spectra show a wide variation depending on their electronic structures.
The U~4$f$ core-level spectra of UB$_2$ consist of a sharp main line with an asymmetric shape, having a tail toward higher binding energies.
In addition to this main line, there is a satellite structure about 5 to 7 eV below the main line but with a very low intensity.
As for UFeGa$_5$ and UPtGa$_5$, their core-level spectra also consist of asymmetric main line and a satellite on the high-binding-energy side.
It is shown that both spectra have very similar spectral line shapes.
Since the core-level spectra are basically dominated by the local electronic structures around the uranium site, it is quite reasonable for these core-level spectra to be similar to each other.
Although their spectral line shapes are considerably different from that of UB$_2$, the essential structures of those spectra (i.e., asymmetric main line and satellite) are similar.

Meanwhile, the core-level spectrum of the localized compound UPd$_3$ has very complex spectral line shape.
Since the spectral line shape is very different from the full-multiplet calculation of the U~4$f$ core-level spectrum of the tetravalent uranium ion\cite{Okada}, it is clear that there are final state screening effects even in a localized 5$f$ compound.
Both the U~4$f_{7/2}$ and U~4$f_{5/2}$ spectra consist of the predominant main line with a relatively broad and symmetric line shape and the broad and intense satellite structure with about 5 to 6~eV higher binding energies of the main line.
In addition to this main line and the satellite structure, a weak shoulder was observed on the low-binding-energy side of the main line, as indicated by the arrow in the figure.
This is the first observation of such a fine peak structure in the core-level spectra of uranium compounds.
This is due to the higher energy resolution in the present experiment ($\Delta E$=120~meV) than in the previous ones ($\Delta E \sim$1~eV).
Hereafter, we refer to this structure as the low-binding-energy shoulder.
The energy separation between the main line and the low-binding-energy shoulder is about 2~eV, which is larger than that of the major multiplet splittings of the multiplet calculation of the U~4$f$ core-level spectrum\cite{Okada}.
Therefore, it is considered that this low-binding-energy shoulder originates from screened final states rather than from multiplet effects.
The existence of this low-binding-energy shoulder suggests that an analysis based on two peaks (main line and satellite) is insufficient to account for the behavior of the core-level spectra.

%-----------------------------------------------------------------------------------------
\begin{figure}
\begin{center}
\includegraphics[scale=0.45]{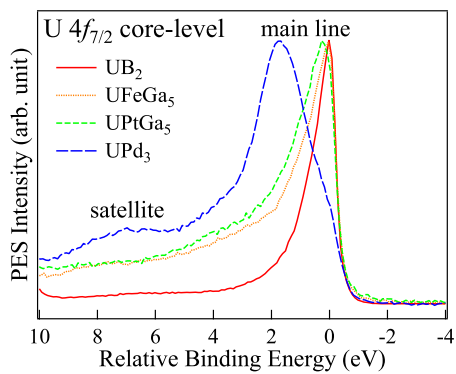}
\caption{(Color-online)
Comparison of the U~4$f_{7/2}$ core-level spectra of UB$_2$, UFeGa$_5$, UPtGa$_5$, and UPd$_3$ taken at $h\nu=800$~eV.
These spectra are aligned so as to match their onsets, and shown on the relative binding energy scale after background subtraction.
}
\label{U4f_comp}
\end{center}
\end{figure}
%-----------------------------------------------------------------------------------------
To see details of these spectral line shapes, we have compared the core-level spectra.
Figure~\ref{U4f_comp} shows the comparison of the U~4$f_{7/2}$ spectra of UB$_2$, UFeGa$_5$, UPtGa$_5$, and UPd$_3$.
These spectra are aligned so as to match their onsets, and shown in relative binding energy.
It is shown that the spectra of the itinerant compounds show some systematic behaviors.
The core-level spectrum of UB$_2$ shows a very sharp and almost single-peak structure, which is close to the core-level spectra of simple metals.
This suggests that the electron correlation effect is weak in this compound.
On the other hand, the core-level spectra of UFeGa$_5$ and UPtGa$_5$ are considerably different from that of UB$_2$.
%%%%%%%%%%%%%%%%%%%%%%%%%%%%%%%%%%%%%%%%%%%%%%
The main lines of UFeGa$_5$ and UPtGa$_5$ are broader than that of UB$_2$ only on the high-binding-energy side.
In addition, the main line of UPtGa$_5$ is slightly broader than that of UFeGa$_5$, although their peak top positions are slightly different.
Since these spectra match well in the energy region of $E_{\rm B}<0$~eV, these changes are not due to a simple broadening or a shift of the entire spectra, but to changes in asymmetry.
Thus, the degree of asymmetry is largest in UPtGa$_5$ and smallest in UB$_2$.
%%%%%%%%%%%%%%%%%%%%%%%%%%%%%%%%%%%%%%%%%%%%%%
Meanwhile, the main line of UPd$_3$ shows a relatively symmetric and broad line shape.
We consider the origin of the main line asymmetry.
The core-level spectra of metallic compounds show asymmetric line shapes due to electron-hole pair creation in the vicinity of $E_{\rm F}$ through a core electron emission process.
This asymmetric line shape of core-level spectra was formulated by Doniach and \u{S}unji\'c, and higher densities of states at $E_{\rm F}$ are expected to cause a more asymmetric line shape\cite{DS}.
In the present case, the main line is more asymmetric in compounds with a larger electronic specific heat coefficient, which is basically proportional to the density of states at $E_{\rm F}$.
Therefore, this difference in the main line asymmetry may originate from the difference in density of states at $E_{\rm F}$ among these four compounds.
%%%%%%%%%%%%%%%%%%%%%%%%%%%%%%%%%%%%%%%%%%%%%%
Meanwhile, this rule does not hold in general since there is another satellite on the lower-binding-energy side of the main line, as in the case of UPd$_2$Al$_3$ shown in the next subsection.
%%%%%%%%%%%%%%%%%%%%%%%%%%%%%%%%%%%%%%%%%%%%%%
In addition to the main line asymmetry, it is shown that the satellite intensity is more enhanced in UPd$_3$ than in other itinerant compounds.
Furthermore, it is stronger in UFeGa$_5$ and UPtGa$_5$ than in UB$_2$, among the itinerant compounds.
The origin of the satellite has been considered as a poorly screened satellite ($f^n$ final state in Fig. \ref{coreprocess}), and it has been conjectured that a weakly hybridized compound shows a higher satellite intensity.
If this is the case, the higher satellite intensities in the core-level spectra of UFeGa$_5$ and UPtGa$_5$ suggest that 5$f$ electrons in these compounds are not as strongly hybridized as that in UB$_2$.
We discuss the behaviors of each satellite in detail in the discussion subsection.

Thus, it is shown that itinerant compounds and localized compounds have very different core-level spectral line shapes.
In itinerant compounds, the spectra consist of an asymmetric main line and a weak satellite on the high-binding-energy side.
The satellite intensity is higher for weakly hybridized and hence more localized compounds.
On the other hand, localized compounds show a relatively symmetric main line with a shoulder structure on its lower-binding-energy side.
In addition, the satellite intensity is higher than the intensities of itinerant compounds.
They are the basis for the understanding of other uranium compounds.

%-----------------------------------------------------------------------------------------
\subsection{Heavy-Fermion superconductors}
%-----------------------------------------------------------------------------------------
\begin{figure}
\begin{center}
\includegraphics[scale=0.45]{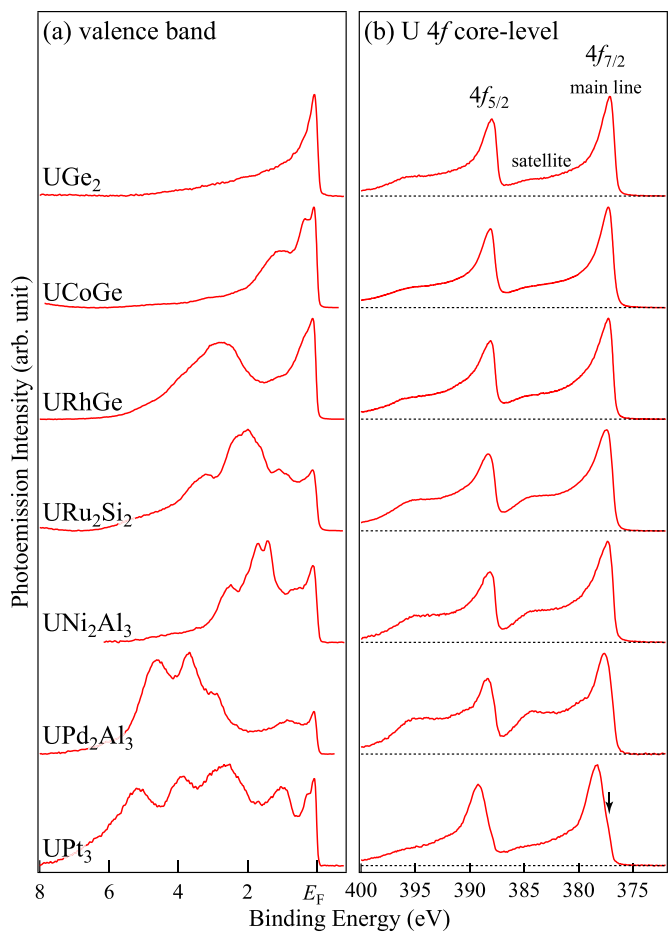}
\caption{(Color-online)
Valence-band spectra (a) and U~4$f$ core-level spectra (b) of UGe$_2$, UCoGe, URhGe, URu$_2$Si$_2$, UNi$_2$Al$_3$, UPd$_2$Al$_3$,~and UPt$_3$ taken at $h\nu=800$~eV.
The lines at the bottom of the spectra show the zero line.
}
\label{HF}
\end{center}
\end{figure}
%-----------------------------------------------------------------------------------------
Next, we discuss the valence-band and core-level spectra of heavy Fermion uranium compounds.
In these compounds, the nature of U~5$f$ states is not yet generally understood as they are in the itinerant and localized compounds shown in the previous subsection.
Therefore, it is important to determine their electronic structures.
Figure~\ref{HF}(a) shows the valence-band spectra of the heavy Fermion superconductors UGe$_2$, UCoGe, URhGe, URu$_2$Si$_2$, UNi$_2$Al$_3$, UPd$_2$Al$_3$, and UPt$_3$.
In these spectra, a relatively sharp peak was observed in the vicinity of $E_{\rm F}$ while complex multiple peaks were observed on the high-binding-energy side.
These are assigned to contributions of the U~5$f$ bands and the transition metal $d$ bands, respectively.
%###########################################################################################
The transition metal $d$ states are centered at approximately $E_{\rm B}=1.5-4$~eV, and they seem to have relatively smaller contributions to $E_{\rm F}$ than U~5$f$ states.
%###########################################################################################
These valence band spectra are different from the spectrum of the localized compound UPd$_3$, suggesting rather itinerant natures of the U~5$f$ states in these compounds.
However, U~5$f$-originated peaks are sharper than the peaks of the itinerant compounds such as UB$_2$ and UFeGa$_5$.
Therefore, the U~5$f$ states form narrow quasi-particle bands near $E_{\rm F}$.

Figure~\ref{HF}(b) shows the U~4$f$ core-level spectra of UGe$_2$, UCoGe, URhGe, URu$_2$Si$_2$, UNi$_2$Al$_3$, UPd$_2$Al$_3$, and UPt$_3$.
These spectra consist of a main line and a satellite as has been observed in other uranium compounds.
Except for UPt$_3$, they show a rather asymmetric main line shape.
Their overall spectral line shapes are similar to those of the itinerant compounds UB$_2$ and U$T$Ga$_5$ rather than that of UPd$_3$, but their satellite intensities are higher in heavy Fermion compounds.
This suggests that U~5$f$ electronic states are not as strongly hybridized as in these itinerant compounds.
The satellite intensity is the highest in UPd$_2$Al$_3$ among the measured nine compounds presented in this paper.
For the U~4$f$ core-level spectrum of UPt$_3$, the main line shows a relatively symmetric line shape, and there is a small shoulder structure on the lower-binding-energy side, as indicated by the arrow in the figure.
The overall feature of the core-level spectrum is similar to that of UPd$_3$ rather than to those of the itinerant and other heavy Fermion compounds.

\subsection{Discussion}
%-----------------------------------------------------------------------------------------
\begin{figure}
\begin{center}
\includegraphics[scale=0.45]{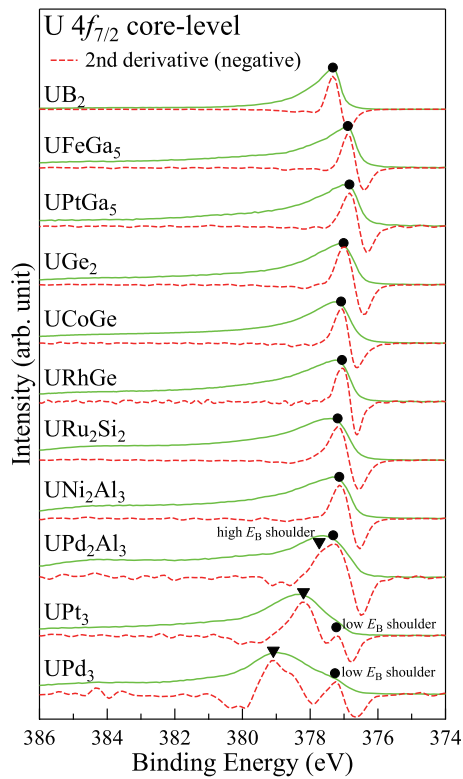}
\caption{(Color-online)
U~4$f_{7/2}$ core-level spectra of UFeGa$_5$, UPtGa$_5$, UGe$_2$, UCoGe, URhGe, URu$_2$Si$_2$, UNi$_2$Al$_3$, UPd$_2$Al$_3$, UPt$_3$, and UPd$_3$ taken at $h\nu=800$~eV, and their negative second derivatives.
Filled circles and filled reversed triangles represent peak positions derived from their second derivatives.
The position of the low-binding-energy shoulder of UPt$_3$ is indicated by each arrow head.}
\label{diff}
\end{center}
\end{figure}
%-----------------------------------------------------------------------------------------
In the preceding subsections, we have shown the behavior of the core-level spectra of various uranium compounds.
In this subsection, we summarize the behavior of the core-level spectra, and attempt to understand the relationship between the core-level spectral line shape and its electronic structure.
It has been shown that the core-level spectra of uranium compounds mainly consist of a dominant asymmetric main line on the low-binding-energy side and a broad satellite structure located at about 6-7~eV higher binding energies than the main line.
In addition to these structures, the main lines of UPd$_3$ and UPt$_3$ are accompanied by a shoulder peak on the lower-binding-energy side.

First, we discuss the behavior of the main line and shoulder peak structures.
To determine their peak positions, we have taken their second derivatives.
Peaks in the negative second derivative correspond to the peak positions in the raw spectrum.
Figure~\ref{diff} shows the U~4$f_{7/2}$ core-level spectra and their negative second derivatives of UB$_2$, UFeGa$_5$, UPtGa$_5$, UGe$_2$, UCoGe, URhGe, URu$_2$Si$_2$, UNi$_2$Al$_3$, UPd$_2$Al$_3$, UPt$_3$, and UPd$_3$.
%This order is the same as that of the magnitude of specific heat coefficient except for UPd$_3$ which is the most localized compound.
For UB$_2$, UFeGa$_5$, UPtGa$_5$, UGe$_2$, UCoGe, URhGe, URu$_2$Si$_2$, and UNi$_2$Al$_3$, the main lines show a rather simple single-peak structure with an asymmetric line shape.
On the other hand, the main lines of UPd$_2$Al$_3$, UPt$_3$, and UPd$_3$ are different from the main lines of these compounds.
In the main line of UPd$_3$, a shoulder peak is located at about 2~eV above the main line.
A similar low-binding-energy shoulder was observed in UPt$_3$ as well and its energy separation from the main line is about 1~eV.
The main line of UPd$_2$Al$_3$ does not show a clear double-peak structure, but it is broader than those of itinerant 5$f$ compounds, and seems to consist of double peaks with a very small energy separation.
These results suggest that the main line consists of two main components, and their intensities and energy separations vary depending on the compound.
Since UPt$_3$ and UPd$_2$Al$_3$ have larger electronic specific heat coefficients than the other uranium compounds, it is expected that the hybridizations between U~5$f$ and ligand states are relatively weak compared with those in the itinerant or other heavy Fermion compounds.
In particular, UPt$_3$ has a very large electronic specific heat coefficient, suggesting that the electron correlation effect is quite strong and therefore its electronic structure is considered to be close to the localized limit.
Therefore, the peak on the high-binding-energy side is dominant in localized or weakly hybridized compounds, while the other peak on the lower-binding-energy side is dominant in the itinerant compounds.
As hybridization the between 5$f$ and ligand states becomes strong, the peak on the high-binding-energy side moves towards lower binding energies, and finally merges with the peak on the lower-binding-energy side.
%Here, we note that the main line of UPd$_3$ has further fine structure at around $E_{\rm B} \sim$ 378.6~eV.
%The energy separation from the main line is about 0.45~eV, and this may be due to the multiplet effects.

%-----------------------------------------------------------------------------------------
\begin{figure}
\begin{center}
\includegraphics[scale=0.5]{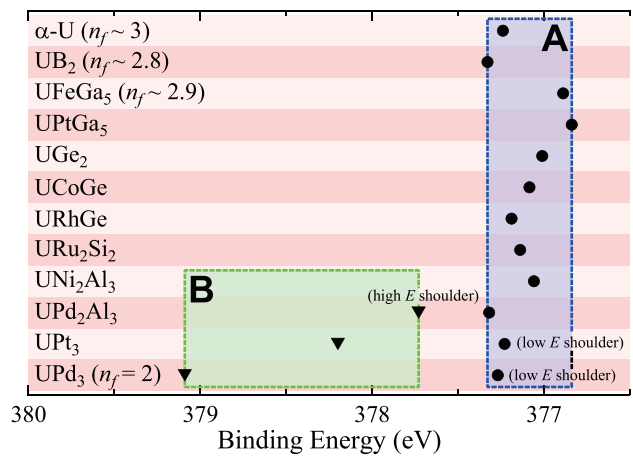}
\caption{(Color-online)
Main line energy position of the U~4$f_{7/2}$ core-level spectra together with the low-binding-energy shoulders of UPt$_3$ and UPd$_3$ and the high-binding-energy shoulder of UPd$_2$Al$_3$.
The main line positions of $\alpha$-U, UB$_2$, UFeGa$_5$, UPtGa$_5$, UGe$_2$, UCoGe, URhGe, URu$_2$Si$_2$, UNi$_2$Al$_3$, and UPd$_2$Al$_3$ as well as the low-binding-energy shoulders of UPt$_3$ and UPd$_3$ are distributed in the energy range of 376.84-377.33~eV, as shown in region A.
}
\label{peakpos}
\end{center}
\end{figure}
%-----------------------------------------------------------------------------------------
Figure~\ref{peakpos} shows a summary of the binding energies of the U~4$f_{7/2}$ main lines estimated from their second derivatives.
The peak position of $\alpha$-U is cited from ref. \cite{alphaU}.
The energy positions of the low-binding-energy shoulders of UPd$_3$ and UPt$_3$, and the high-binding-energy shoulder of UPd$_2$Al$_3$ are also shown in the figure.
It is shown that the binding energies of the U~4$f_{7/2}$ main line of eight of the ten compounds including the itinerant and some heavy Fermion compounds fall into the energy window of $E_{\rm B}$ = 376.84-377.33~eV ($\Delta E \sim$0.5~eV), as indicated by region A.
In addition, the low-binding-energy shoulders observed in UPd$_3$ and UPt$_3$ are located within this energy region.
This seems to suggest that the main lines of $\alpha$-U through UPd$_2$Al$_3$ as well as the low-binding-energy shoulders of UPt$_3$ and UPd$_3$ have the same final state character.
Here, we consider the final state character in this energy region by examining the cases of $\alpha$-U, UB$_2$, and UFeGa$_5$.
The number of $f$ electrons in $\alpha$-U was estimated to be $n_f$=3 from the electron energy loss (EELS) experiments\cite{alphaU2}.
Meanwhile, since a very good agreement between the band-structure calculation and ARPES spectra has been obtained for UB$_2$\cite{UB2_ARPES} and UFeGa$_5$\cite{UFeGa5_ARPES}, we estimate the numbers of 5$f$ electrons in the ground states by band-structure calculation.
The numbers of 5$f$ electrons estimated from the band-structure calculations are 2.82 for UB$_2$ and 2.89 for UFeGa$_5$.
Thus, we assume that the numbers of 5$f$ electrons in $\alpha$-U, UB$_2$, and UFeGa$_5$ are close to three.
In the final state of the photoemission process, a strong hybridization between 5$f$ states and ligand states of neighboring 5$f$ states in these compounds leads to the screening of core-hole potential by the transfer of ligand electrons into lowered 5$f$ states.
This results in a dominant 5$f^4$ configuration character of the final state.
Hence, the peaks in region A are assigned to the 5$f^4$ dominated final state.
Here, note that a stronger hybridization between 5$f$ and ligand states leads to more mixed nature of pure final-state configurations ($f^n$) in each peak than those observed in rare-earth 4$f$ compounds.
%Therefore, those final states are strongly mixed, and the assignments mentioned above are nominal ones.

%----------check----------check----------check----------check----------check----------
On the other hand, the main lines of UPd$_3$ and UPt$_3$, and the high-binding-energy shoulder of UPd$_2$Al$_3$ are located on the higher-binding-energy side of region A.
Although they have different binding energies, their positions show a systematic behavior.
As 5$f$ electrons become well hybridized (in the order of UPd$_3$, UPt$_3$, and UPd$_2$Al$_3$), peaks move toward lower binding energies.
This suggests that the peaks in region B have the same final-state character, and that this shift corresponds to different hybridization strengths in these compounds.
A resonant photoemission experiment on UPd$_{3-x}$Pt$_x$ alloys shows that their valence-band 5$f$ spectra also show a double-peak structure with different final states\cite{UPd3_RPES}.
The peak on the higher-binding-energy side is dominant in UPd$_3$ while the peak near $E_{\rm F}$ becomes dominant as alloys approach UPt$_3$.
In addition, the former peak shifts toward lower binding energies by about 0.4~eV, while the latter peak stays at $E_{\rm F}$ in going from $x$=0 (UPd$_3$) to $x$=3 (UPt$_3$).
This behavior is very similar to that of the main line and low-binding-energy shoulders of UPd$_3$ and UPt$_3$, suggesting that the main line and the low-binding-energy shoulder originate from different final states.
Here, we consider the character of the peaks in region B by considering the case of UPd$_3$.
The initial electronic configuration of UPd$_3$ is localized 5$f^2$, and the most plausible interpretation of its core-level spectrum is that the main line and the satellite on the high-binding-energy side are contributions mainly of the well-screened 5$f^3$ final state and poorly screened 5$f^2$ final state, respectively.
Hence, the main line of UPd$_3$ can be assigned to the final state dominated by the 5$f^3$ configuration.
A similar splitting of the main line has been observed in the Ru~3$d$ core-level spectra of ruthenates, and it has been interpreted as a different screening number of $4d$ states in the final state\cite{ruthenates}.
This is very similar to the present interpretation.
%----------check----------check----------check----------check----------check----------

Accordingly, we have found that there are two main components in the vicinity of the main line: one on the high-binding-energy side corresponding to the 5$f^3$ dominant final state and one on the lower-binding-energy side corresponding to the 5$f^4$ final state.
In itinerant compounds with a nearly 5$f^3$ configuration, the 5$f^4$ final state becomes dominant.
As the 5$f$ electronic states become more localized, the 5$f^3$ final state appears on the high-binding-energy side.
In localized or nearly localized compounds, this 5$f^3$ final state becomes dominant.
These behaviors of the main line can be utilized to understand how 5$f$ electrons are localized or itinerant in uranium compounds.
The main lines of UGe$_2$, UCoGe, URhGe, URu$_2$Si$_2$, and UNi$_2$Al$_3$ are relatively simple single structures located in region A.
This suggests that 5$f$ electrons in these compounds have an essentially itinerant character.

%-----------------------------------------------------------------------------------------
Next, we discuss the behavior of the satellite.
It has been considered that the satellite is a poorly screened final state, and it is conjectured that less hybridized compounds show higher satellite intensities.
Such a tendency is observed for the compounds measured in our present study.
For example, the satellite intensity of the localized compound UPd$_3$ is much higher than those of the itinerant compounds and most of the heavy Fermion compounds.
However, there are also exceptions.
For example, the satellite intensity of UPd$_2$Al$_3$ is higher than or comparable to that of UPd$_3$, although UPd$_3$ is the most localized compound.
Hence, the above generalization is not a strict rule, although there is such a trend in general.
Nanba and Okada suggested that the line shape of non-$5f$ ligand bands may has a strong effect on the line shape of core-level spectra since the hybridization between $f$ and ligand states is strong in 5$f$ compounds\cite{Nanba}.
Therefore, a realistic model calculation is required to understand the exact behavior of satellite peak structures.

%-----------------------------------------------------------------------------------------
Although its microscopic mechanism was not fully clarified in the present study, the existence of a satellite in the core-level spectra of heavy Fermion compounds suggests the correlated nature of 5$f$ electrons in these compounds.
For UPd$_2$Al$_3$, quasi-particle bands were observed in the vicinity of $E_{\rm F}$ by ARPES experiments, and the overall band structure was well explained by the band-structure calculation assuming all 5$f$ electrons as itinerant\cite{UPd2Al3_ARPES}.
This suggests that 5$f$ electrons have a very itinerant character in this compound.
However, the existence of a strong satellite in its core-level spectrum suggests that 5$f$ electrons are in a much correlated state.
%----------------put-some-more-descriptions-----------------------------------------------
In particular, the similarity of the core-level spectra between UPd$_3$ and UPt$_3$ implies that the 5$f$ 
electronic state of UPt$_3$ is very close to the localized limit.
This is consistent with the fact that very large electron masses have been observed in nearly localized low-$T_{\rm K}$ compounds\cite{Settai}.
%----------------put-some-more-descriptions-----------------------------------------------
Finally, we comment on the valence states of uranium atoms in these compounds.
The dominant 5$f^4$ final state in the core-level spectra of UGe$_2$, UCoGe, URhGe, URu$_2$Si$_2$, UNi$_2$Al$_3$, and UPd$_2$Al$_3$ suggests the dominance of the 5$f^3$ initial state configuration in the ground state.
Meanwhile, the finite satellite intensity suggests that there should also be a 5$f^2$ configuration in the initial state.
Therefore, the numbers of 5$f$ electrons in these compounds in the ground states are close to but less than three.
On the other hand, the number of 5$f$ electrons in UPt$_3$ and UPd$_3$ in the ground state should be close to two rather than to three since the 5$f^3$ final state is dominant in its core-level spectrum.
%-----------------------------------------------------------------------------------------

\section{Conclusions}

We have measured the valence-band and core-level spectra of UB$_2$, UFeGa$_5$, UPtGa$_5$, UGe$_2$, UCoGe, URhGe, URu$_2$Si$_2$, UNi$_2$Al$_3$, UPd$_2$Al$_3$, UPt$_3$, and UPd$_3$.
The results are summarized as follows:

(i) The overall structure of the core-level spectra of uranium compounds consists of the dominant main line on the low-binding-energy side and the satellite structure on the high-binding-energy side.

(ii) The main lines consist basically of two components with an energy separation of about or 2~eV or less.
The peak on the high-binding-energy side is dominant in localized or weakly hybridized compounds whereas the peak on the lower-binding-energy side is dominant in itinerant or strongly hybridized compounds.
Consideration based on their valence states suggests that the former is the 5$f^3$ dominant final state whereas the latter is the 5$f^4$ dominant final state.

(iii) In general, the satellite intensity is higher in more localized or weakly hybridized compounds.

(iv) The main lines of UGe$_2$, UCoGe, URhGe, URu$_2$Si$_2$, and UNi$_2$Al$_3$ are relatively simple-single peak structures located at lower binding energies, suggesting that 5$f$ electrons in these compounds are well hybridized.

(v) Although the core-level spectrum of UPd$_2$Al$_3$ shows similarity to those of itinerant uranium compounds, the existence of a very strong satellite and a high-binding-energy shoulder of the main line suggests that U~5$f$ electrons have a very well correlated nature due to the weaker hybridization in UPd$_2$Al$_3$ than in the other heavy Fermions.

(vi) The core-level spectrum of UPt$_3$ is similar to that of UPd$_3$ rather than to those of itinerant compounds, suggesting that the electronic structure of UPt$_3$ is very close to the localized limit.

(vii) The numbers of 5$f$ electrons in UGe$_2$, UCoGe, URhGe, URu$_2$Si$_2$, UNi$_2$Al$_3$, and UPd$_2$Al$_3$ are close to but less than to three.
On the other hand, that of UPt$_3$ should be close to two rather than three.

\section*{Acknowledgments}
We would like to thank K.~Okada, Y.~Nanba, and J.~Joyce for stimulating discussion and comments.
The present work was financially supported by a Grant-in-Aid for Scientific Research from the Ministry of Education, Culture, Sports, Science and Technology, Japan under contract No. 21740271, by a Grant-in-Aid for Scientific Research on Innovative Areas "Heavy Electrons" (No. 20102003) from The Ministry of Education, Culture, Sports, Science and Technology, Japan, and by the Shorei Kenkyuu Funds from Hyogo Science and Technology Association.

%-----------------------------------------------------------------------------------------

\end{document}